\documentclass[12pt]{article}

\usepackage{epsfig}
\usepackage{amsmath}
\usepackage{amssymb}
\usepackage{axodraw}

\providecommand{\beqa}{\begin{eqnarray}}
\providecommand{\eeqa}{\end{eqnarray}}
\providecommand{\ap}{{\alpha^\prime}}
\providecommand{\Dta}{{\overline{D3}}}

\def\cH{{\cal{H}}}
\def\cM{{\cal{M}}}
\def\cO{{\cal{O}}}
\def\cS{{\cal{S}}}

\begin{document}

\begin{titlepage}
\begin{flushright}
UMD-PP-04-014\\
hep-th/0312309\\
\today
\end{flushright}

\vspace{1cm}
\begin{center}
\baselineskip25pt {\Large\bf On the Bekenstein-Hawking Entropy,
Non-Commutative Branes and Logarithmic Corrections}

\end{center}
\vspace{1cm}
\begin{center}
\baselineskip12pt {Axel Krause\footnote{E-mail: {\tt
akrause@fas.harvard.edu}, now at Jefferson Physical Laboratory,
Harvard University, Cambridge, MA 02138, USA}}

\vspace{1cm}

{\it Center for String and Particle Theory,}\\[1.8mm]
{\it Department of Physics, University of Maryland}\\[1.8mm]
{\it College Park, MD 20742, USA}

\vspace{0.3cm}
\end{center}
\vspace*{\fill}

\begin{abstract}
We extend earlier work on the origin of the Bekenstein-Hawking
entropy to higher-dimensional spacetimes. The mechanism of
counting states is shown to work for all spacetimes associated
with a Euclidean doublet $(E_1,M_1)+(E_2,M_2)$ of
electric-magnetic dual brane pairs of type II string-theory or
M-theory wrapping the spacetime's event horizon plus the complete
internal compactification space. Non-Commutativity on the brane
worldvolume enters the derivation of the Bekenstein-Hawking
entropy in a natural way. Moreover, a logarithmic entropy
correction with prefactor $1/2$ is derived.
\end{abstract}


\vspace*{\fill}

\end{titlepage}

\section{Introduction}

In the past few years M/string-theory made considerable progress
towards the stabilization of its multitude of moduli arising from
compactification and therefore towards predictivity
\cite{GVW}-\cite{BCK}. Particularly interesting for addressing the
real world is heterotic M-theory \cite{HW1}-\cite{LOW} with its
flux compactifications \cite{HetFlux1}-\cite{HetFlux3}. The latter
offer an intriguing rationale for why and how grand unified
theories have to be combined with gravity. Due to its peculiar
blend of classical and quantum physics \cite{HW2}, it is clear
that eventually knowledge of the full quantum heterotic M-theory
becomes intimately tied to its promising phenomenology. It is
therefore also from a very pragmatic phenomenological point of
view of direct interest to obtain the elusive non-perturbative
formulation of M-theory which would require first of all to
identify its fundamental microscopic states.

Unfortunately, experimentally, we are still far away from testing
M-theory directly. In order to unravel the mysteries of M-theory
we therefore have to look for other, necessarily theoretical,
clues. Probably the best guidance in this respect comes from the
Bekenstein-Hawking entropy (BH-entropy) \cite{BHE1}-\cite{BHE7}.
Because of its universal applicability it seems to capture a
generic feature of the underlying microscopic theory of quantum
gravity. In this paper we would like to propose, following earlier
work, \cite{K11}-\cite{K14}, a possible set of microscopic
chain-like states for the non-perturbative regime where the
string-coupling constant $g_s$ becomes of order one.

We will base our argumentation on a microscopic mechanism to count
states leading to the BH-entropy and its leading logarithmic
correction. Along the line of reasoning which will involve
electric-magnetic dual Euclidean brane pairs, we will see that
non-commutativity on the brane worldvolume fits in naturally,
suggesting a non-commutative event horizon. Let us mention that,
also within the ``membrane paradigm'' approach (see
e.g.~\cite{BHM1},\cite{BHM2}), it was recently pointed out in
\cite{Li} that the stretched horizon of a black hole should be
thought of as a non-commutative membrane, suggesting as well a
non-commutative event horizon. The results which we will present
here will generalize the analysis of \cite{K11} in that they apply
also to the BH-entropy of higher-dimensional $d>4$ spacetimes
while \cite{K11} was devoted to the study of the BH-entropy of
$d=4$ spacetimes. For some earlier other interesting ideas trying
to understand the BH-entropy from string- or M-theory see
\cite{BHS1}-\cite{BHS9}.

\section{BH-Entropy and Electric-Magnetic Dual Pairs}

\subsection{Type II String-Theory Case}

Let us start with type II string-theory on a D=10 geometric
background $\cM^{1,d-1}\times\cM^{10-d}$ described by a metric
($2\le d\le 10$)
\beqa
ds^2 = g_{\mu\nu}^{(1,d-1)}(x) dx^\mu dx^\nu + g_{mn}^{(10-d)}(y)
dy^m dy^n
\label{Metric}
\eeqa
with Lorentzian signature. The $d$-dimensional external
non-compact space $\cM^{1,d-1}$ is the one whose BH-entropy we are
interested in while the internal space $\cM^{10-d}$ is taken to be
compact with certain factorization properties as we will explain
below. More specifically, because we are interested in an external
spacetime with non-zero BH-entropy, let $\cM^{1,d-1}$ possess a
$(d-1)$-dimensional future event horizon $\cH^+$. Consider a
$(d-1)$-dimensional spacelike hypersurface $\Sigma$ in
$\cM^{1,d-1}$ with one boundary at spatial infinity $i_0$ and
another boundary, denoted $\cH^{d-2}$, on $\cH^+$. For spacetimes
$\cM^{1,d-1}$ describing black holes, $\cH^{d-2}$ is commonly
referred to as the ``boundary'' of the black hole and its area as
the ``area of the horizon''.

Next, let us take a pair of mutually orthogonal (with respect to
the metric given in (\ref{Metric})) Euclidean electric-magnetic
dual type II branes
\beqa
(E_1,M_1) \in \{(\text{Dp},\text{D(6-p)}),(\text{F1},\text{NS5})\}
\; .
\label{Set1}
\eeqa
Without exhibiting them explicitly, we understand that this set of
dual type II branes includes as well all possible pairs in which
any brane (including for short also the fundamental string F1) is
replaced by its charge-reversed anti-brane. As the orthogonal
Euclidean $E_1$ and $M_1$ together can cover an 8-dimensional
submanifold, they possess the right dimensionality so that we can
wrap the pair $(E_1,M_1)$ around the complete $\cH^{d-2} \times
\cM^{10-d}$. In case the (Euclidean) dimensions of $E_1$ and $M_1$
do not happen to coincide with $d-2$ resp.~$10-d$ (in either
order) we would require $\cM^{10-d}$ resp.~$\cH^{d-2}$ to
factorize appropriately. For instance a torus compactification
$\cM^{10-d}=T^{10-d}$ would satisfy this requirement. In the
special case of a Schwarzschild black hole in uncompactified
10-dimensional spacetime one would have to include the dual
Euclidean pair $(\text{D7},\text{D(-1)})$ as well.

For reasons which will soon become clear, we will wrap a second
pair of mutually orthogonal Euclidean electric-magnetic dual
branes
\beqa
(E_2,M_2)\in \{(\text{Dq},\text{D(6-q)}),(\text{F1},\text{NS5})\}
\eeqa
around $\cH^{d-2} \times \cM^{10-d}$. Here $(E_1,M_1)$ and
$(E_2,M_2)$ can be chosen independently as far as the mechanism of
counting states will be concerned. However, the choice of
$(E_1,M_1)$ and $(E_2,M_2)$ has to be compatible with the metric
background (\ref{Metric}). For instance, a Schwarzschild black
hole background geometry which is charge neutral, will require a
second antibrane pair $(E_2,M_2) \equiv (\bar{E}_1,\bar{M}_1)$ to
neutralize the charges of the first pair.

In passing let us mention that wrapping branes around the horizon
is very reminiscent of the ``membrane paradigm'' in which the
event horizon of a $d=4$ black hole (or rather its ``stretched
horizon'') is conceived as an effective membrane (however not a
fundamental one like in M/string-theory) which enjoys some
intriguing non-relativistic properties \cite{BHM}. One might
therefore view part of the current approach also as an embedding
of this idea into M/string-theory where the role of the effective
membrane is played by fundamental M/string-theory branes. Note,
however, that while the effective membrane is Lorentzian and has a
built in time direction, this is not the case for the Euclidean
branes. The inherently small lifetime of a Euclidean brane, which
is of order $\Delta t \sim \sqrt{\ap}/c$, gets however infinitely
dilated by the time dilatation between the event horizon and any
exterior observer, as we will discuss below.

The reason for introducing the dual pairs is that it will allow a
useful rewriting of the BH-entropy, associated with the
$(d-2)$-dimensional area of $\cH^{d-2}$, as we will see next. Our
goal is to express this BH-entropy exclusively in terms of
string-theory entities. Since each pair $(E_i,M_i);\,i=1,2$ covers
the space $\cH^{d-2} \times \cM^{10-d}$, we can write for the
compactification volume in both cases
\begin{equation}
vol(\cM^{10-d}) = \frac{vol(E_i)vol(M_i)}{vol(\cH^{d-2})} \;\; ,
\;\;\; i=1,2 \; .
\end{equation}
The effective $d$-dimensional Newton Constant can therefore be
expressed as
\begin{equation}
G_d = \frac{G_{10}}{vol(\cM^{10-d})} = \frac{(2\pi)^6\ap^4
g_s^2}{8}\times \frac{vol(\cH^{d-2})}{vol(E_i) \, vol(M_i)}
\;\;,\;\;\; i=1,2
\end{equation}
where $\ap$ denotes the Regge slope.

The significance of why we have chosen to use dual branes lies in
the fact that the product of their tensions satisfies the
generalized Dirac quantization condition
\begin{equation}
\tau_{E_i}\tau_{M_i} = \frac{1}{(2\pi)^6\ap^4g_s^2} \; .
\end{equation}
which allows us to express the inverse of the Newton Constant as
\begin{equation}
\frac{1}{G_d} = \frac{8\,\big(\tau_{E_i}vol(E_i)\big)\,
\big(\tau_{M_i}vol(M_i)\big)}{vol(\cH^{d-2})}
\;\;, \;\;\; i=1,2 \; .
\end{equation}
Hence, the BH-entropy of the spacetime $\cM^{1,d-1}$, associated
with the area of $\cH^{d-2}$, can be written purely in terms of
the respective Euclidean Nambu-Goto actions $S_{E_i},S_{M_i}$ for
$E_i$ and $M_i$ as
\begin{equation}
{\cal S}_{BH} = \frac{vol(\cH^{d-2})}{4G_d} = 2S_{E_i}S_{M_i} \;\;
, \;\;\; i=1,2 \; ,
\label{BH1}
\end{equation}
where $S_{E_i} = \tau_{E_i} vol(E_i)$ and similarly for the
magnetic branes.

The fact that we considered two dual pairs instead of just one
becomes important now. Namely, it allows us to get rid of the
prefactor 2 and write the BH-entropy instead as a sum over both
dual pairs
\begin{equation}
{\cal S}_{BH} = \sum_{i=1,2} S_{E_i}S_{M_i} \; .
\label{BH2}
\end{equation}
For the mechanism to microscopically derive $\cS_{BH}$ by counting
states, which we will present later, it will be important that we
can replace the prefactor 2 by the sum at the expense of
introducing two dual pairs instead of just one. If we had to use
only one dual pair then the microscopic derivation of $\cS_{BH}$
would be off by precisely this factor 2. Moreover, taking two dual
pairs instead of just one also makes sense from the point of view
that a charge-neutral Schwarzschild black hole wouldn't be
compatible with just one dual pair. At least one other anti-brane
pair is needed to dispose of the long-range U(1) RR- or NS-fields
of the branes.

Let us finally comment on the stability of the Euclidean brane
system. A Euclidean brane would exist for just a very short
time-interval, given essentially by the string-time $\Delta t \sim
\sqrt{\ap}/c$, if put into a flat spacetime background. In
contrast, the Euclidean branes considered here, are located on an
event horizon. The infinite redshift with which an exterior
observer sees the horizon, implies, at the classical level, an
infinite time dilatation, rendering the Euclidean branes
classically stable to any such observer. It is worthwhile
emphasizing that the entities which characterize an event
horizon's thermodynamics -- and in particular its entropy -- such
as the mass, the angular momentum and charge of the interior
region of spacetime enclosed by it, are also measured away from
the horizon, in the exterior asymptotic regime. We should
therefore analyze the system of Euclidean branes not from an
observer's point of view who is located on these branes but from
an exterior observer's point of view for whom they appear stable
due to the infinite time dilatation \cite{K12}.

\subsection{M-Theory Case}

Let us now try to obtain an analogous expression for the
BH-entropy in M-theory. In M-theory we have a unique dual brane
pair $(\text{M2},\text{M5})$ and essentially the same reasoning
goes through as in the type II case. We will start with a D=11
spacetime $\cM^{1,d-1} \times \cM^{11-d}$ whose geometry describes
a compactification from 11 to $d$ dimensions $(2\le d\le 11)$
\beqa
ds^2 = g_{\mu\nu}^{(1,d-1)}(x)dx^\mu dx^\nu
+ g_{mn}^{(11-d)}(y)dy^m dy^n \; .
\label{MMetric}
\eeqa
The external non-compact $d$-dimensional spacetime $\cM^{1,d-1}$
is the one whose BH-entropy we are interested in. We therefore
assume that it has a non-trivial future event horizon $\cH^+$. The
$(d-2)$-dimensional intersection of $\cH^+$ with a spacelike
hypersurface $\Sigma$, coming in from spatial infinity $i_0$, is
again denoted $\cH^{d-2}$. The volume of $\cH^{d-2}$ is known as
the ``area of the horizon''.

The pair $(\text{M2},\text{M5})$ of a Euclidean M2 and M5 brane
which are mutually orthogonal in the metric (\ref{MMetric}),
covers a 9-dimensional spacelike submanifold. We will let the pair
wrap $\cH^{d-2} \times \cM^{11-d}$. Except for $d=5$ and $d=8$, we
would have to assume that the metric on either $\cM^{11-d}$ or
$\cH^{d-2}$ factorizes appropriately into a direct product. For
$d=9,10,11$ both M2 and M5 would have to wrap $\cH^{d-2}$ whose
metric must then exhibit a corresponding direct product structure.
At first, this seems to exclude the Schwarzschild black holes in
$d=9,10,11$ dimensions from consideration as in these cases
$\cH^{d-2}$ is spherical and spheres don't factorize. We had,
however, seen before that the $d=9,10$ cases can be covered by the
richer type II brane description which e.g.~also allows for a dual
$(\text{D7},\text{D(-1)})$ Euclidean brane pair to cover the
$d=10$ situation. So it is really the $d=11$ Schwarzschild case
which cannot be adressed. It might be possible to include it as
well by invoking the less understood M9-brane but will not pursued
further here.

For M-theory compactifications the compactification volume can
then be expressed as
\beqa
vol(\cM^{11-d})=\frac{vol(\text{M2})vol(\text{M5})}{vol(\cH^{d-2})}
\eeqa
such that the effective $d$-dimensional Newton Constant becomes
\beqa
G_d = \frac{G_{11}}{vol(\cM^{11-d})}
= \frac{(2\pi)^7l_{11}^9}{8}\times\frac{vol(\cH^{d-2})}
{vol(\text{M2})vol(\text{M5})}
\eeqa
where $l_{11}$ denotes the 11-dimensional Planck-length. The
important property of the dual brane pair is that the product of
their tensions satisfies
\beqa
\tau_{\text{M2}}\tau_{\text{M5}} = \frac{1}{(2\pi)^7l_{11}^9} \; .
\eeqa
This allows to write the inverse of $G_d$ as
\beqa
\frac{1}{G_d} =
\frac{8\left(\tau_{\text{M2}}vol(\text{M2})\right)
\left(\tau_{\text{M5}}vol(\text{M5})\right)}
{vol(\cH^{d-2})} \; .
\eeqa
The $d$-dimensional BH-entropy associated with the spacetime
$\cM^{1,d-1}$ can therefore be expressed as
\beqa
\cS_{BH} = \frac{vol(\cH^{d-2})}{4G_d}
= 2S_{\text{M2}}S_{\text{M5}}
\label{BH3}
\eeqa
where $S_{\text{M2}},S_{\text{M5}}$ are the respective Nambu-Goto
actions of the Euclidean M2 and M5.

For a second dual brane pair $(\text{M2},\text{M5})$ wrapped
around $\cH^{d-2} \times \cM^{11-d}$ independently of the first
pair (for the following expression of the BH-entropy, the M2's
resp.~M5's of the two pairs don't have to wrap necessarily the
same submanifolds) one would arrive at the same result
(\ref{BH3}). Hence one obtains, by employing two
$(\text{M2},\text{M5})$ pairs, also in the M-theory case the
result
\beqa
\cS_{BH} = \sum_{i=1,2} S_{\text{M2},i}S_{\text{M5},i} \; ,
\label{BH11}
\eeqa
which expresses the $d$-dimensional BH-entropy exclusively in
terms of the branes' Nambu-Goto actions. Again, since the
Nambu-Goto action does not recognize the difference between a
brane and an anti-brane we will understand subsequently that each
M2 or M5 could also be replaced by its anti-brane partner.

\section{Cell Structure and Non-Commutativity}

Now that we have found an expression for the BH-entropy in terms
of the Nambu-Goto actions of two dual brane pairs, our aim will be
to propose a suitable set of microstates capable of explaining the
entropy by counting the states of a microcanonical ensemble. For
this we will need one more ingredient to which we will turn now.
When dealing with a Euclidean brane, it is more natural to treat
its Euclidean ``time'' and space dimensions not differently in
contrast to the case of a Lorentzian brane where the Lorentzian
signature leads to such a distinction. Consequently, one is led to
interpret the tension of for instance a $(p+1)$-dimensional
Euclidean Dp-brane not as its ``mass'' per unit spatial $p$-volume
but instead as the inverse of a $(p+1)$-dimensional volume unit
$v_{\text{Dp}}$. On any of the Euclidean branes introduced so far
we will therefore have a volume unit $v_E$ resp.~$v_M$ given by
the inverse of the brane's tension
\beqa
v_E = \frac{1}{\tau_E} \; ,
\qquad v_M = \frac{1}{\tau_M} \; .
\label{MinVol}
\eeqa

Such an elementary volume unit on the brane's worldvolume can be
naturally understood if the brane's worldvolume would be
considered being non-commutative. Taking the simplest
non-commutativity arising from string-theory
\cite{SWNC1},\cite{SWNC2}
\beqa
[X^i,X^j]=2i\epsilon^{ij}l^2
\eeqa
for the worldvolume coordinates $X^i$ of a Euclidean Dp-brane, one
derives the uncertainty relation
\beqa
\Delta X^i \Delta X^j \ge l^2
\eeqa
for the coordinates $X^i$. From this follows directly the ``brane
worldvolume uncertainty principle'' \cite{CHK}
\beqa
\Delta X^1\hdots\Delta X^{p+1} \ge l^{p+1} \; .
\label{WVUP}
\eeqa
Indeed in \cite{CHK} such an uncertainty principle was shown to
arise in string field theory for all branes (including F1, NS5,
Dp, M2, M5) by using S- and T-dualities. The result was that the
smallest allowed volume $l^{p+1}$ in (\ref{WVUP}) is always given
(up to factors of $\cO(1)$) through the tension of the respective
brane
\beqa
l^{p+1}\simeq \frac{1}{\tau_{\text{Dp}}}
\eeqa
for all Dp-branes and similarly for F1, NS5, M2, M5. From the
perspective of a brane with non-commutative worldvolume (which
arises in string field theory even in the absence of a background
magnetic flux or Neveu-Schwarz $B$-field along the brane
\cite{CHK}), it is therefore clear that $v_E,v_M$ in
(\ref{MinVol}) represent the {\em smallest volume unit} which is
allowed by the ``worldvolume uncertainty principle''. For the
special case of the fundamental string this just states that
$2\pi\ap = 1/\tau_{\text{F1}}$ constitutes a smallest volume unit
resp.~that the string-length $l_s$ constitutes a smallest length
-- a familiar result which has been argued for based on string
scattering amplitudes, worldsheet conformal invariance and other
arguments \cite{MinString1},\cite{MinString2}.

With this interpretation of the tension of a Euclidean brane, its
Nambu-Goto action adopts a new meaning. Namely, quite analogous to
the decomposition of phase space into cells of size $h$ in quantum
mechanics, we are led here to think of the brane worldvolume as a
lattice composed out of a certain number of cells with volume
$v_E$ resp.~$v_M$. It is then precisely the Nambu-Goto action of
the brane, $S_E$ resp.~$S_M$, which counts how many such cells,
$N_E$ resp.~$N_M$, the brane contains
\begin{equation}
N_E = \frac{vol(E)}{v_E} = \tau_E vol(E) = S_E
\label{NGC}
\end{equation}
and similarly for the magnetic dual component
\begin{equation}
N_M = S_M \; .
\label{NGC2}
\end{equation}
Because of the orthogonality of $E$ and $M$, the cells of the dual
pair $(E,M)$ will be of size
\begin{equation}
V_{cell} = v_E v_M \; .
\end{equation}
Hence, each dual pair $(E,M)$ contains
\begin{equation}
\frac{vol(E)vol(M)}{V_{cell}} =
\frac{vol(E)}{v_E}\frac{vol(M)}{v_M}
= N_E N_M
\end{equation}
cells. Since the product of the tensions of two dual branes is
independent of the specifically chosen dual branes, $V_{cell}$ is
the same for all dual pairs. Two dual pairs $(E_1,M_1)$ and
$(E_2,M_2)$, both covering the same volume, will therefore contain
\beqa
\frac{vol(E_1)vol(M_1)}{V_{cell}} +
\frac{vol(E_2)vol(M_2)}{V_{cell}}
= \sum_{i=1,2} N_{E_i} N_{M_i}
= N
\label{NOC}
\eeqa
cells. By virtue of (\ref{BH2}) resp.~(\ref{BH11}) plus
(\ref{NGC}), (\ref{NGC2}) together with (\ref{NOC}), this implies
that the $d$-dimensional BH-entropy associated with $\cM^{1,d-1}$
simply becomes an integer
\beqa
\cS_{BH} = \sum_{i=1,2} S_{E_i} S_{M_i} = \sum_{i=1,2} N_{E_i}
N_{M_i} = N \in 2\mathbf{N}\; ,
\label{BHInt}
\eeqa
with {\em $N$ the total number of cells} contained in the combined
worldvolume of $(E_1,M_1)+(E_2,M_2)$. This result is valid at
sufficiently large $N$ where $N\gg \Delta$ and we can neglect
expected but unknown microscopic quantum corrections $\Delta$ and
set $N+\Delta \simeq N$. Tiny as these small quantum corrections
may be, they will generically shift the value of the corrected
expression $N+\Delta$ away from being an integer. We will in the
following not consider $\Delta$ further and work in the
$N\gg\Delta$ regime. Notice that $N$ has to be even because both
$(E_1,M_1)$ and $(E_2,M_2)$ cover the same volume which implies
the equality
\begin{equation}
S_{E_1}S_{M_1} = \frac{vol(E_1)vol(M_1)}{V_{cell}} =
\frac{vol(E_2)vol(M_2)}{V_{cell}}
= S_{E_2}S_{M_2}
\end{equation}
and leads to $N_{E_1}N_{M_1}=N_{E_2}N_{M_2}$.

\setcounter{figure}{0}
\begin{figure}[t]
\begin{center}
\begin{picture}(260,120)(0,20)
\Text(10,112)[]{$1$} \Line(0,20)(0,100) \Line(20,20)(20,100)
\Line(0,100)(20,100) \Line(0,86)(20,86) \Line(0,72)(20,72)
\Line(0,34)(20,34) \Line(0,20)(20,20) \Text(10,93)[]{$1$}
\Text(10,79)[]{$2$} \Text(10,63)[]{$\vdots$}
\Text(10,51)[]{$\vdots$} \Text(10,27)[]{$N$} \Line(10,51)(66,79)

\Text(70,112)[]{$2$} \Line(60,20)(60,100) \Line(80,20)(80,100)
\Line(60,100)(80,100) \Line(60,86)(80,86) \Line(60,72)(80,72)
\Line(60,34)(80,34) \Line(60,20)(80,20) \Text(70,93)[]{$1$}
\Text(70,79)[]{$2$} \Text(70,63)[]{$\vdots$}
\Text(70,51)[]{$\vdots$} \Text(70,27)[]{$N$} \Line(74,79)(126,79)

\Text(130,112)[]{$3$} \Line(120,20)(120,100)
\Line(140,20)(140,100) \Line(120,100)(140,100)
\Line(120,86)(140,86) \Line(120,72)(140,72) \Line(120,34)(140,34)
\Line(120,20)(140,20) \Text(130,93)[]{$1$} \Text(130,79)[]{$2$}
\Text(130,63)[]{$\vdots$} \Text(130,51)[]{$\vdots$}
\Text(130,27)[]{$N$} \Line(134,79)(165,30)

\Text(180,60)[]{$\hdots$}

\Text(230,112)[]{$N$} \Line(220,20)(220,100)
\Line(240,20)(240,100) \Line(220,100)(240,100)
\Line(220,86)(240,86) \Line(220,72)(240,72) \Line(220,34)(240,34)
\Line(220,20)(240,20) \Text(230,93)[]{$1$} \Text(230,79)[]{$2$}
\Text(230,63)[]{$\vdots$} \Text(230,51)[]{$\vdots$}
\Text(230,27)[]{$N$} \Line(195,70)(230,50)
\end{picture}
\caption{An open chain consisting of a sequence of $N-1$ links.
Each link can start and end on any of the $N$ cells of the joint
dual pairs' lattice represented schematically by each of the
columns.}
\label{OpenChains}
\end{center}
\end{figure}
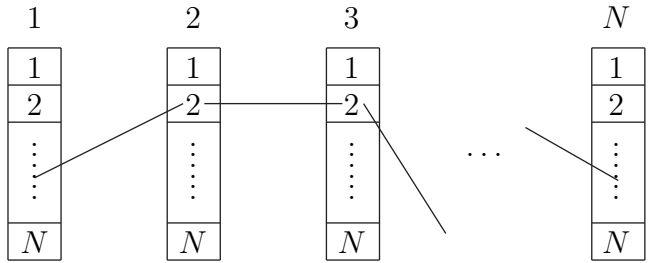
The scaling of the BH-entropy with an integer $N$ at large $N$
shouldn't be a surprise because it is precisely what the
holographic principle \cite{Holo} demands. Holography implies that
the number of fundamental degrees of freedom $N_{\text{dof}}$ of a
system does not scale with the volume but with its area in
Planckian units. In our case the area is the area of the horizon
\beqa
N_{\text{dof}}\sim \frac{vol(\cH^{d-2})}{G_d} \; .
\eeqa
With the BH-entropy itself being proportional to the area of the
horizon, one obtains, via the holographic principle, a scaling of
the BH-entropy with an integer, namely the number of fundamental
degrees of freedom
\beqa
\cS_{BH} \sim N_{\text{dof}} \; .
\eeqa
{\em Holography therefore demands that we identify} $N_{\text{dof}}
\sim N$ {\em and consequently interpret the cells, or rather one
degree of freedom per cell, as the $N$ fundamental degrees of
freedom}.

Let us note that in the limit where the string coupling constant
$g_s\rightarrow 0$ goes to zero and we understand branes as smooth
hypersurfaces, we won't see the discrete cell structure. The
reason is that the smallest allowed discrete volume
$V_{cell}=1/(\tau_E\tau_M)\propto g_s^2\rightarrow 0$ goes to zero
in this limit and the involved worldvolumes become quasi-smooth.
However, in the non-perturbative regime where $g_s\simeq 1$,
$V_{cell}$ is finite and the discrete brane worldvolumes become
visible.

\section{Derivation of BH-Entropy and Logarithmic Correction}

\begin{figure}[t]
\begin{center}
\begin{picture}(260,160)(0,10)
\Line(30,35)(30,115)
\Line(50,35)(50,115)
\Text(40,127)[]{$2$}
\Line(30,115)(50,115)
\Text(40,108)[]{$1$}
\Line(30,101)(50,101)
\Text(40,94)[]{$2$}
\Line(30,87)(50,87)
\Text(40,78)[]{$\vdots$}
\Text(40,66)[]{$\vdots$}
\Line(30,49)(50,49)
\Text(40,42)[]{$N$}
\Line(30,35)(50,35)

\Line(60,70)(60,150)
\Line(80,70)(80,150)
\Text(70,162)[]{$3$}
\Line(60,150)(80,150)
\Text(70,143)[]{$1$}
\Line(60,136)(80,136)
\Text(70,129)[]{$2$}
\Line(60,122)(80,122)
\Text(70,113)[]{$\vdots$}
\Text(70,101)[]{$\vdots$}
\Line(60,84)(80,84)
\Text(70,77)[]{$N$}
\Line(60,70)(80,70)

\Line(88,10)(88,90)
\Line(108,10)(108,90)
\Text(98,102)[]{$1$}
\Line(88,90)(108,90)
\Text(98,83)[]{$1$}
\Line(88,76)(108,76)
\Text(98,69)[]{$2$}
\Line(88,62)(108,62)
\Text(98,53)[]{$\vdots$}
\Text(98,41)[]{$\vdots$}
\Line(88,24)(108,24)
\Text(98,17)[]{$N$}
\Line(88,10)(108,10)

\Line(120,120)(73,129)
\Line(67,129)(40,60)
\Line(40,60)(98,52)
\Line(98,52)(162,28)
\Line(162,28)(220,63)
\Line(220,63)(190,110)
\Line(190,110)(140,135)
\Text(130,130)[]{$\hdots$}

\Line(152,10)(152,90)
\Line(172,10)(172,90)
\Text(162,102)[]{$N$}
\Line(152,90)(172,90)
\Text(162,83)[]{$1$}
\Line(152,76)(172,76)
\Text(162,69)[]{$2$}
\Line(152,62)(172,62)
\Text(162,53)[]{$\vdots$}
\Text(162,41)[]{$\vdots$}
\Line(152,24)(172,24)
\Text(162,17)[]{$N$}
\Line(152,10)(172,10)

\Line(180,70)(180,150)
\Line(200,70)(200,150)
\Text(190,162)[]{$N-2$}
\Line(180,150)(200,150)
\Text(190,143)[]{$1$}
\Line(180,136)(200,136)
\Text(190,129)[]{$2$}
\Line(180,122)(200,122)
\Text(190,113)[]{$\vdots$}
\Text(190,101)[]{$\vdots$}
\Line(180,84)(200,84)
\Text(190,77)[]{$N$}
\Line(180,70)(200,70)

\Line(210,35)(210,115)
\Line(230,35)(230,115)
\Text(220,127)[]{$N-1$}
\Line(210,115)(230,115)
\Text(220,109)[]{$1$}
\Line(210,102)(230,102)
\Text(220,95)[]{$2$}
\Line(210,88)(230,88)
\Text(220,79)[]{$\vdots$}
\Text(220,67)[]{$\vdots$}
\Line(210,49)(230,49)
\Text(220,42)[]{$N$}
\Line(210,35)(230,35)
\end{picture}
\caption{A closed chain possesses one more link than an open chain
which is necessary to close the chain by connecting the last and
the first link on the same cell. The closed chain gives the same
number of different states as the open chain.}
\label{ClosedChains}
\end{center}
\end{figure}
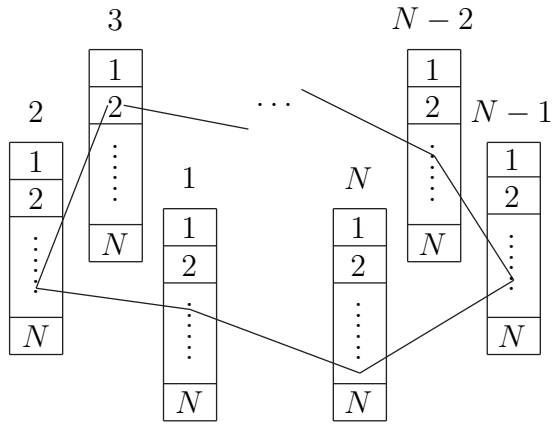

Next, we want to propose a set of microscopic states whose entropy
should, in a microcanonical ensemble description, account for the
BH-entropy and ideally also for its logarithmic correction. For
this purpose we will consider on the combined $(E_1,M_1) +
(E_2,M_2)$ worldvolume, taken as a lattice of $N$ cells, open
chains built out of $N-1$ successive links where each link
connects two arbitrary cells (see fig.\ref{OpenChains}). As each
link will be allowed to start and end on any of the cells with
same probability, the number of all such chain configurations is
clearly $N^N$. Besides the open chains there is a similar but
topologically different class of chains which possesses the same
number of configurations, $N^N$. These are the closed chains made
out of $N$ links where the last link connects back to the first
link (see fig.\ref{ClosedChains}).
As far as the state counting is concerned they lead to the same
results as the open chains and might therefore be considered as an
alternative set of microstates until further selection criteria
are found.

We could have called the counting of the different chain
configurations so far ``classical'' because it considered all
cells as being distinguishable. The cells, which we identified as
the fundamental degrees of freedom via holography, should however
at the quantum level rather be regarded as bosonic degrees of
freedom and therefore considered being indistinguishable. We will
see that this quantum feature leads to the correct counting of
states. As well-known from statistical mechanics, one can easily
account for this quantum bosonic symmetry at large $N$ by dividing
the classical number of configurations through the
Gibbs-correction factor $N!$. Therefore, {\em
quantum-mechanically} we obtain a number of
\begin{equation}
\Omega(N)=\frac{N^N}{N!}
\end{equation}
different open or closed chain states.

By assuming that all chains with the same length $N-1$ (open)
resp.~$N$ (closed) will possess the same energy, we can now
determine in a microcanonical ensemble description the chain
entropy. It is given by
\begin{equation}
{\cal S}_{chain} = \ln\Omega(N)
\end{equation}
and can be evaluated in the large $N$ limit (as appropriate for an
area of $\cH^{d-2}$ of macroscopic size) by using Stirling's
series
\beqa
N! = \sqrt{2\pi N} N^N e^{-N} \left(
1+\frac{1}{12N}+\cO\Big(\frac{1}{N^2}\Big)\right)
\eeqa
to approximate $\ln (N!)$. The result is
\begin{equation}
\cS_{chain} = N - \frac{1}{2}\ln N - \ln \sqrt{2\pi} -
\frac{1}{12N}
+{\cal O}\Big(\frac{1}{N^2}\Big)
\end{equation}
which by virtue of the identification (\ref{BHInt}) becomes
\beqa
\cS_{chain} = \cS_{BH} - \frac{1}{2}\ln\cS_{BH} - \ln \sqrt{2\pi}
-\frac{1}{12\cS_{BH}} + \cO\Big(\frac{1}{\cS_{BH}^2}\Big) \; .
\eeqa

Thus indeed the chain entropy matches at leading order the
BH-entropy and in addition at subleading order gives the expected
logarithmic correction including the precise numerical prefactor
(there has been a debate in the literature over whether the
prefactor in front of the logarithm should be $1/2$ or $3/2$.
While initially a factor $3/2$ had been universally favored
\cite{Drei1},\cite{Drei2} arguments have been put forward more
recently favoring $1/2$, see \cite{LC2a}-\cite{LC2d}. The
discussion is still ongoing but it is certainly clear that the
prefactor depends on the ensemble chosen and will change when one
leaves the microcanonical ensemble description and uses the less
fundamental canonical ensemble \cite{Debate1},\cite{Debate2}).

We can therefore conclude that the proposed microscopic chain
states for the non-perturbative $g_s\simeq \cO(1)$ regime allow
for a general mechanism of counting states with the correct
reproduction of the BH-entropy and its leading logarithmic
correction not only for 4-dimensional \cite{K11} but also, as
demonstrated in this paper, for $d$-dimensional spacetimes
possessing event horizons.

\section{Final Comments}

Let us briefly address the situation where one treats the
Euclidean branes not as probe branes, as we have done it here, but
includes their backreaction on the spacetime geometry. First steps
in this direction for the 4-dimensional Schwarzschild black hole
have been undertaken in \cite{K12}. It is clear that the uncharged
$d$-dimensional hyperspherically symmetric
Schwarzschild-Tangherlini black hole has to be associated with
branes where the second doublet, $(E_2,M_2) \equiv ({\bar
E}_1,{\bar M}_1)$ consists of the anti-branes of the first doublet
in order to be compatible with an uncharged configuration. This
also fits because both the Schwarzschild-Tangherlini black hole
and the brane anti-brane configuration break all supersymmetry.
Moreover, because the Schwarzschild-Tangherlini black hole is a
non-dilatonic black hole, one should employ in the type II case a
self-dual $(D3,D3)+(\Dta,\Dta)$ doublet, as the $D3$-brane is the
only non-dilatonic type II brane.

Let us finally return to the cell-volume. As mentioned before, the
cell-volume $V_{cell}$ on each of the two dual pairs $(E_i,M_i)$
is given, due to the orthogonality of $E_i$ and $M_i$, by the
product of their minimal volumes
\beqa
V_{cell} = v_{E_i}v_{M_i} = \frac{1}{\tau_{E_i}\tau_{M_i}}
= \left\{
  \begin{array}{ll}
  (2\pi)^6 \ap^4 g_s^2 & \quad (D=10) \\
  (2\pi)^7 l_{11}^9    & \quad (D=11)
  \end{array}
  \right. \; .
\eeqa
Restoring the constants $\hbar$ and $c$ and noticing that
$G_d\hbar/c^3$ has length-dimension $L^{d-2}$, this result can be
written as
\beqa
V_{cell}
= \left\{
  \begin{array}{ll}
  8G_{10}\frac{\hbar}{c^3} & \quad (D=10) \\
  8G_{11}\frac{\hbar}{c^3} & \quad (D=11)
  \end{array}
  \right. \; .
\label{CellGN}
\eeqa
It shows first that the D=10/11 Newton Constant acquires a
geometric meaning in terms of the cell volume and second that a
non-vanishing cell-volume is a {\em quantum effect} which vanishes
in the classical limit where $\hbar\rightarrow 0$. This agrees
also with the understanding of the cell-volume in terms of a
non-commutative structure on the brane worldvolume which likewise
results from a promotion of the classical coordinates $x^i$ to
non-commuting quantum operators $X^i$. Moreover, we see that when
gravity becomes weak, i.e.~when $G_{10,11}$ becomes small, the
cell volume shrinks and we end up with the ordinary smooth
hypersurface description of branes which we expect from
perturbative string-theory in this regime. On the other hand, it
is clear that when gravity becomes strong, i.e.~when $G_{10,11}$
becomes large, the discrete cell structure should show up
prominently and hence signals a significant deviation from
ordinary string-theory in the non-perturbative regime.

Given the relations (\ref{CellGN}) in D=10/11 dimensions, one
readily obtains formulae for the effective $d$-dimensional Newton
Constant $G_d$ (and effective Planck-length $l_d$ defined through
$l_d^{d-2}=G_d \frac{\hbar}{c^3}$) which are purely geometrical.
These are
\beqa
G_d\frac{\hbar}{c^3}
= \frac{G_{10}}{vol(\cM^{10-d})} \frac{\hbar}{c^3}
= \frac{V_{cell}}{8vol(\cM^{10-d})} \qquad (D=10)
\label{CellGd10}
\eeqa
for the 10-dimensional type II case and
\beqa
G_d\frac{\hbar}{c^3}
= \frac{G_{11}}{vol(\cM^{11-d})} \frac{\hbar}{c^3}
= \frac{V_{cell}}{8vol(\cM^{11-d})} \qquad (D=11)
\label{CellGd11}
\eeqa
for the 11-dimensional M-theory case. The size of the effective
$d$-dimensional Newton Constant appears as the ratio of the cell
volume to the compactification volume.

\bigskip
\noindent {\large \bf Acknowledgements}\\[2ex]
This work has been supported by the National Science Foundation
under Grant Number PHY-0099544.

\newcommand{\zpc}[3]{{\sl Z.Phys.} {\bf C\,#1} (#2) #3}
\newcommand{\npb}[3]{{\sl Nucl.Phys.} {\bf B\,#1} (#2) #3}
\newcommand{\npps}[3]{{\sl Nucl.Phys.(Proc.Suppl.)} {\bf #1} (#2) #3}
\newcommand{\plb}[3]{{\sl Phys.Lett.} {\bf B\,#1} (#2) #3}
\newcommand{\prd}[3]{{\sl Phys.Rev.} {\bf D\,#1} (#2) #3}
\newcommand{\prb}[3]{{\sl Phys.Rev.} {\bf B\,#1} (#2) #3}
\newcommand{\pr}[3]{{\sl Phys.Rev.} {\bf #1} (#2) #3}
\newcommand{\prl}[3]{{\sl Phys.Rev.Lett.} {\bf #1} (#2) #3}
\newcommand{\prsla}[3]{{\sl Proc.Roy.Soc.Lond.} {\bf A\,#1} (#2) #3}
\newcommand{\jhep}[3]{{\sl JHEP} {\bf #1} (#2) #3}
\newcommand{\cqg}[3]{{\sl Class.Quant.Grav.} {\bf #1} (#2) #3}
\newcommand{\prep}[3]{{\sl Phys.Rep.} {\bf #1} (#2) #3}
\newcommand{\fp}[3]{{\sl Fortschr.Phys.} {\bf #1} (#2) #3}
\newcommand{\nc}[3]{{\sl Nuovo Cimento} {\bf #1} (#2) #3}
\newcommand{\nca}[3]{{\sl Nuovo Cimento} {\bf A\,#1} (#2) #3}
\newcommand{\lnc}[3]{{\sl Lett.~Nuovo Cimento} {\bf #1} (#2) #3}
\newcommand{\ijmpa}[3]{{\sl Int.J.Mod.Phys.} {\bf A\,#1} (#2) #3}
\newcommand{\ijmpd}[3]{{\sl Int.J.Mod.Phys.} {\bf D\,#1} (#2) #3}
\newcommand{\rmp}[3]{{\sl Rev. Mod. Phys.} {\bf #1} (#2) #3}
\newcommand{\ptp}[3]{{\sl Prog.Theor.Phys.} {\bf #1} (#2) #3}
\newcommand{\sjnp}[3]{{\sl Sov.J.Nucl.Phys.} {\bf #1} (#2) #3}
\newcommand{\sjpn}[3]{{\sl Sov.J.Particles\& Nuclei} {\bf #1} (#2) #3}
\newcommand{\splir}[3]{{\sl Sov.Phys.Leb.Inst.Rep.} {\bf #1} (#2) #3}
\newcommand{\tmf}[3]{{\sl Teor.Mat.Fiz.} {\bf #1} (#2) #3}
\newcommand{\jcp}[3]{{\sl J.Comp.Phys.} {\bf #1} (#2) #3}
\newcommand{\cpc}[3]{{\sl Comp.Phys.Commun.} {\bf #1} (#2) #3}
\newcommand{\mpla}[3]{{\sl Mod.Phys.Lett.} {\bf A\,#1} (#2) #3}
\newcommand{\cmp}[3]{{\sl Comm.Math.Phys.} {\bf #1} (#2) #3}
\newcommand{\jmp}[3]{{\sl J.Math.Phys.} {\bf #1} (#2) #3}
\newcommand{\pa}[3]{{\sl Physica} {\bf A\,#1} (#2) #3}
\newcommand{\nim}[3]{{\sl Nucl.Instr.Meth.} {\bf #1} (#2) #3}
\newcommand{\el}[3]{{\sl Europhysics Letters} {\bf #1} (#2) #3}
\newcommand{\aop}[3]{{\sl Ann.~of Phys.} {\bf #1} (#2) #3}
\newcommand{\jetp}[3]{{\sl JETP} {\bf #1} (#2) #3}
\newcommand{\jetpl}[3]{{\sl JETP Lett.} {\bf #1} (#2) #3}
\newcommand{\acpp}[3]{{\sl Acta Physica Polonica} {\bf #1} (#2) #3}
\newcommand{\sci}[3]{{\sl Science} {\bf #1} (#2) #3}
\newcommand{\nat}[3]{{\sl Nature} {\bf #1} (#2) #3}
\newcommand{\pram}[3]{{\sl Pramana} {\bf #1} (#2) #3}
\newcommand{\hepth}[1]{{\tt hep-th/}{\tt #1}}
\newcommand{\hepph}[1]{{\tt hep-ph/}{\tt #1}}
\newcommand{\grqc}[1]{{\tt gr-qc/}{\tt #1}}
\newcommand{\astroph}[1]{{\tt astro-ph/}{\tt #1}}
\newcommand{\desy}[1]{{\sl DESY-Report~}{#1}}

\bibliographystyle{plain}

\end{document}